\documentclass[prl,aps,amsfonts,amssymb,amsmath,graphicx,showpacs,twocolumn,epsfig,amscd,ulem]{revtex4}

\usepackage{graphicx}
\usepackage{natbib}
\usepackage{amssymb}
\usepackage{amsmath}
\usepackage{verbatim}

\begin{document}

\title{Experimental evidence for the intricate free-energy landscape for a soft glassy system}

\author{Sara Jabbari-Farouji $^{1}$, Gerard H.\ Wegdam $^{1}$ and
Daniel Bonn$^{1,2}$}

\affiliation{$^{1}$Van der Waals-Zeeman Institute, University of
Amsterdam, 1018XE Amsterdam, the Netherlands }

\affiliation{$^{2}$Laboratoire de Physique Statistique, UMR 8550
CNRS associated with the University of Paris 6 and Paris 7, Ecole
Normale Sup\'erieure, 75231 Paris Cedex 05, France }

\date{\today}

\begin{abstract}

In the free energy
landscape picture of glassy systems, the slow dynamics  characteristic of these
systems is believed to be due to the existence of a complicated free-energy landscape with many
 local minima. We show here that for a colloidal glassy material
multiple paths can be taken through the free energy landscape,
that can even lead to different 'final' non-ergodic states at the
late stages of aging. We provide clear experimental evidence
for the distinction of gel and glassy states in the system and show that for a range of colloid concentrations, the transition to
non-ergodicity can occur in either direction (gel or glass), and may be accompanied by 'hesitations' between the two directions. This
 shows that colloidal gels and glasses are merely global free-energy minima in
 the same free energy landscape, and that the paths leading to these minima can indeed be complicated.

\end{abstract}

\maketitle

The main issue in the study of glasses is that from the
point of view of their molecular structure, they closely resemble
liquids. On the other hand, their mechanical properties are much
closer to those of solids: ordinary window glass, for instance,
does not flow on human timescales  \cite{church}. To explain the extremely high
viscosity of glassy systems, it is commonly accepted that the
motion of molecules or particles that constitute the glass are
blocked by the neighboring molecules, who in turn are blocked also
by their neighbors and so on, making it impossible for the system to flow.

Translated in terms of the free energy of the system, the paradigm
for glasses is that of a complicated free energy landscape, with
many local minima of the free energy \cite{landscape,landscape1}.
The blocking of molecules or particles is then equivalent to
saying that they are trapped in a local minimum of the free energy
from which they cannot escape. This is most evident in the "aging"
of such systems \cite{landscape,landscape1}: just after a quench
into the glassy phase, the system still evolves in the sense that
both the mechanical properties and the diffusion coefficient
change in time. The interpretation of this time evolution is that
at early times after its formation, the system is able to access
at least part of the phase space, and can get out of local minima
by thermal activation. However, as time goes on, the system finds
deeper and deeper minima, more difficult to escape from, and
consequently the evolution becomes slower. Because of this, the
system cannot reach thermodynamic equilibrium: it remains
non-ergodic. During this aging, the viscosity increases and the
diffusion coefficient of the particles decreases, emphasizing the
link between the blocking of the particle motion and absence of
flow.

However, although this provides an appealing intuitive picture of
glassy dynamics (or rather, the absence of dynamics), to our
knowledge there is no direct experimental evidence for the
existence of such a complicated free-energy landscape with many
local minima \cite{models}. In this Letter we are able to provide
such evidence by showing that for a soft glassy material multiple
paths can be taken through the free energy landscape, that can
even lead to different 'final' non-ergodic states at the late
stages of aging.

The system we consider is a suspension of anisotropic and charged
colloidal particles suspended in water: Laponite, a synthetic clay
\cite{kroon, Mourchid,glass, Italian, Bonn1, Nicolai,Bellour}. The
study of colloids allowed for a significant contribution to
elucidating the basic physics of glass transition
\cite{HS,attractive glass}. In colloidal systems, as the particle
volume fraction is increased, the particles become increasingly
slower and for even higher volume fractions the glass transition
is encountered. On the other hand, colloidal gels are known to
form at extremely low volume fractions $10^{-4}-10^{-2}$ in the
presence of strong attractions \cite{gel1}. Gelation and the glass
transition have important similarities. Both are ergodic to
non-ergodic transitions that are
 kinetic, rather than thermodynamic in origin, and distinguishing between
 these two types of non-ergodic states
experimentally is a longstanding controversy \cite{Tanaka}. The
experiments reported below provide direct criteria for
distinguishing gels from glasses. This allows us to show that for
a range of Laponite concentrations, two distinctly different
non-ergodic states can result at late times: either the glass or
the gel forms at late times with roughly equal probability. There
is no way to tell beforehand which of the two options will be
taken by the sample, suggesting that there are at least two
metastable minima in the system. In addition, our results show
that the free energy landscape is indeed complicated, since a
number of samples are observed to hesitate between the two options
for a long time and an initial evolution in one of the two
directions can lead to a final state that is the other one.

The formation of non-ergodic states in our system is followed in
time using light scattering. The non-ergodicity parameter
(Debye-Waller factor) deduced from these measurements, (which
quantifies the fraction of frozen-in fluctuations \cite{HS}) falls
onto either of two master curves, showing that two possible routes
towards non-ergodicity exist. Measurement of translational
diffusion and scattered intensity then allows to unambiguously
identify these two distinct non-ergodic states as colloidal gel
and glass.

For the experiments, we prepare Laponite XLG suspensions in
ultrapure water. After stirring for two hours, we filtered using
Millipore $0.8\mu m$ filter units to obtain a reproducible initial
state \cite{glass}. This instant defines the zero of aging time,
$t_{a}=0$. A standard dynamic light scattering ($\lambda = 632.8
nm$) measures the time-averaged intensity correlation functions
$g_{\mbox{\scriptsize t}}(q,t)
=\frac{<I(q,t)I(q,0)>_t}{<I(q,0)>_t^{2}}$,  at scattering wave
vector $q=\frac{4\pi n}{\lambda}\sin (\frac{\theta}{2})$, in which
$\theta=90^{o}$ is the scattering angle. The correlation functions
are measured at a rate depending on the speed of aging of
different Laponite suspensions. The aging time for which the
time-averaged correlation functions are not equal to their
ensemble-averaged values, i.e. their values change from one
position to another in the sample,  defines the
ergodicity-breaking point $t_{eb}$. For aging times $t_a >
t_{eb}$, we calculate the ensemble-averaged electric field
correlation function i.\ e.\ intermediate scattering function
$f(q,t)$ from the time-averaged intensity correlation function
$g_t(q,t)$ and ensemble-averaged intensity $I_E$ measured by
rotating the sample at different heights \cite{kroon,HS}.
\begin{equation} \label{eq:a1}
f(q,t)= 1+ (I_t/I_E)\{[g_t(q,t)-g_t(q,0)+1 ]^{1/2}-1 \}
\end{equation}
\begin{figure}
\includegraphics[scale=0.6]{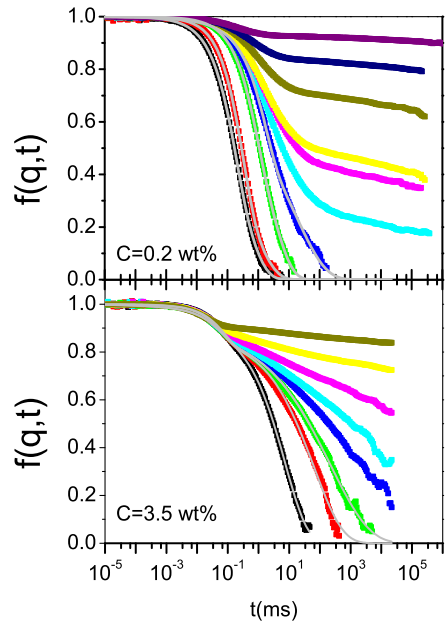}
\caption{Evolution of intermediate scattering function for two
Laponite concentrations, at scattering angle $90^{o}$. The symbols present the
measured correlation functions at increasing waiting times (from
left to right) that are ($t_{a}=1, 49, 71, 80, 85, 89, 91, 99,
112, 141$ days) for $C= 0.2 $ wt\% and ($t_{a}=7, 40, 54, 71, 86,
113, 260, 1356 $ min) for $C=3.5$ wt\%. In both panel, the lines
show the fits with
$A\exp(-t/\tau_{1})+(1-A)\exp(-(t/\tau_{2})^{\beta})$.}\label{fig1}
\end{figure}

 Figure (\ref{fig1}) shows the evolution of intermediate scattering
functions $f(q,t)$ for two different samples. In both cases
correlation functions evolve from an ergodic state to a
non-ergodic state (the correlation function no longer decays to
zero), as the system ages. However, the low-and high-concentration
samples are seen to behave in a distinctly different manner. In
the non-ergodic regime, the aging rate of the system can be
quantified by measuring the time evolution of the non-ergodicity
parameter $f(q,\infty)$, defined as the long-time limit of the
intermediate scattering function.
 Plotting the non-ergodicity
parameter as a function of scaled aging time reveals that its
evolution in a range of samples with different Laponite
concentrations collapses onto two distinct master curves when
plotted (Fig.\ 2a) as a function of reduced aging time $(t_{a}/
t_{eb}-1)$. In the first group of samples (low concentrations) the
non-ergodicity parameter almost reaches
 unity: the colloidal particles are completely blocked, suggesting that they are rigidly held in place, as they would be in a
  gel-like structure. In
  the second group (high concentrations), the non-ergodicity parameter evolves at a slower rate and goes
 to $\approx 0.8$ at late times, indicating that there is still some freedom for the particles to move: the
  hindrance of the particles is only sterical, as it would be in a glass. In fact measuring the non-ergodicity parameter for a
couple of glassy samples until $t_a/t_{eb}=15$, we found that
$f(q,\infty)$ did not exceed 0.835.

\begin{figure}\
\includegraphics[scale=0.6]{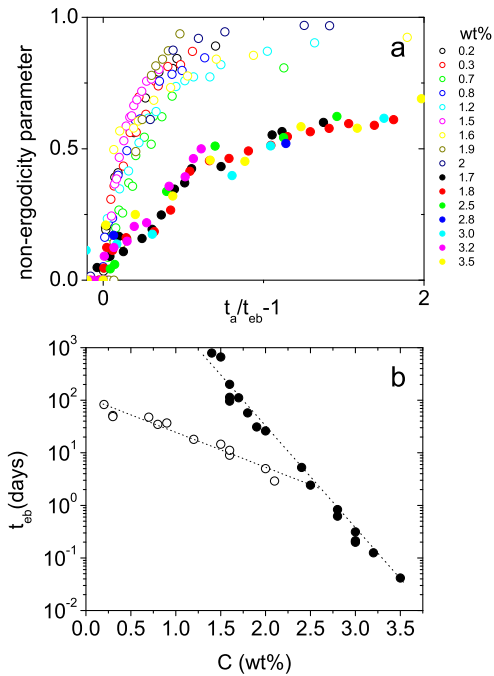}
\caption{ a) The evolution of the non-ergodicity parameter
$f(q,\infty)$ versus reduced aging time $t_a/t_{eb}-1$ for
different Laponite samples. The colloid concentrations are shown
in the legend. The samples can be divided into two groups
according to the evolution of non-ergodicity parameters. In both
figures (and the following ones), the open symbols correspond to
gels, the filled symbols to the glass. b) The ergodicity-breaking
time as a function of concentration.
 }\label{fig2}
 \end{figure}

Perhaps the most striking observation is that for the intermediate
concentrations $1.3 <  C < 2.3 wt\% $, identical samples may
evolve at very different rates, thus having a very different
ergodicity-breaking point. In Fig.\ (\ref{fig2}b), we have plotted
the ergodicity-breaking time of the ensemble of the samples we
have measured as a function of concentration. The samples fall in
two separate groups, and the samples in the intermediate
concentration region  fall in either of the two. This suggests
that the intermediate concentration samples have two options,
either following the same trend as the samples of high
concentration or behaving similarly to the samples of low
concentrations.

\begin{figure}
\includegraphics[scale=0.5]{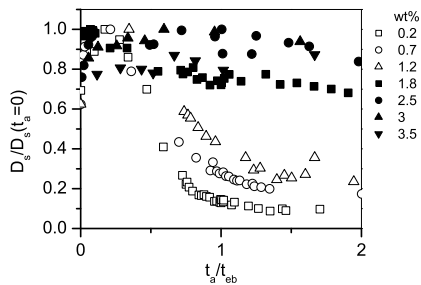}
\caption{a) The evolution of short-time translational diffusion
normalized to its initial value $(t_a \approx 0)$ as a function of
$t_a / t_{eb}$}\label{fig3}
 \end{figure}

\begin{figure}
\includegraphics[scale=0.45]{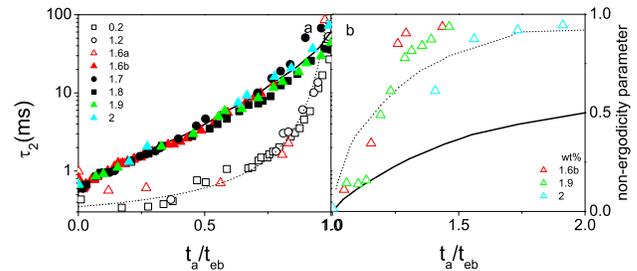}
\caption{ a) Slow relaxation time $\tau_2$  as a function of
scaled aging time b) non-ergodicity parameter as a function of
scaled aging time. These samples behave like a glass at the early
stages of aging and a after $t_w/t_{eb}\approx 1.2$ they evolve
according to the gel line. The filled lines show the glass and
dotted lines show the gel line, obtained by smoothed averaging
over all the samples measured.
 }\label{fig4}
 \end{figure}

The difference between the two states becomes clear if we look at
the rest of the data. Before ergodicity breaking, as can be
observed from the fits shown in  Fig.\ (\ref{fig1}), both the
low-and the high concentration samples can be described  by a sum
of a single exponential and stretched exponential  $A
\exp(-t/\tau_{1})+(1-A)\exp(-(t/\tau_{2})^{\beta})$
 \cite{kroon,Bonn1,Italian}.  $\tau_{1}$ is related to the inverse of the
 short-time diffusion $\tau_1=1/ (D_s q^2)$; as is shown in fig. \ref{fig3},
 the short time diffusion coefficient is almost a constant for the samples
 of high concentration $(C> 2.3 \, wt\%)$; this corresponds to the 'rattling in the cage' motion reported
 earlier for colloidal glasses \cite{Bonn1}.  However, $D_s$ decreases significantly with aging time for low concentrations $(C < 1.4 \, wt\%)$ which is again related to the incorporation of
  the colloidal particles in a gel network \cite{Italian}. In addition, the slow relaxation time  $\tau_2$, is found to grow exponentially with aging time for high concentrations,
in agreement with earlier observations for the glassy state
\cite{Italian,Bonn1}. However, for the gel phase, the relaxation
time increases faster than exponentially (Fig.\ \ref{fig4}a); this
is likely to be related to the formation of small clusters in the
beginning that subsequently aggregate to form a macroscopic
structure (with a large relaxation time) \cite{Nicolai,Italian} as
in diffusion-limited cluster aggregation (DLCA).


\begin{figure}
\includegraphics[scale=1.2]{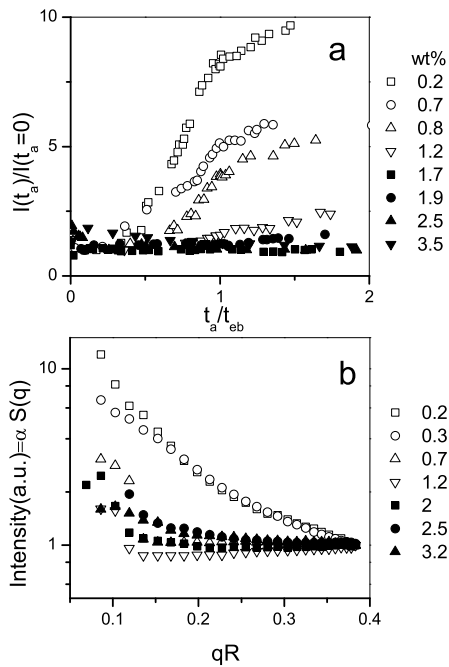}
\caption{a) Scatted intensity at scattering angle $90^{o}$ as a
function reduced aging time. So as to focus on the effect of
aging, we have normalized the intensity to its value at the
beginning of aging. b) The scattered intensity as a function of
dimensionless wave vector $qR$ for several Laponite
concentrations. To be able to compare the q-dependence of
different samples, we have normalized the data to their value at
highest measured $qR=0.38$. As we observe the higher the Laponite
concentration, the more homogenous it becomes. }\label{fig5}
 \end{figure}

The measured static structure factor (Fig.\ \ref{fig5}) indeed
provides two further pieces of evidence for structure formation.
First, for all the low-concentration samples, the scattered
intensity consistently increases with aging time, while for the
high concentration ones it does not evolve much (Fig.\
\ref{fig5}a). The increase of scattered light intensity is usually
attributed to formation a network or clusters of particles
\cite{gel1,gel2,Nicolai}. Second, although the range of wave
vectors is rather limited, measurement of the static structure
factor $S(q)$ for the different samples , a power law behavior for
$S(q)$ appears to be observed for the lowest concentration samples
with an exponent $1.1\pm 0.2$. This exponent is indeed
characteristic of DLCA, in which less compact clusters are formed
than in diffusion-limited aggregation \cite{gel1,gel2}. To the
contrary, an almost flat structure factor is found for the highest
concentrations, indicating homogeneity (Fig.\ (\ref{fig5}b), very
characteristic of a glass \cite{glass}. The noise at low $q$ for
these measurements is probably due to imperfections of the
measurement cell that scatter light at small angles.

Combining all these data, we identify the low-concentration samples
 as colloidal gels and the high-concentration ones as colloidal glasses.
  Intermediate concentrations can be either gels or glasses at late times, with no way of
  telling beforehand how the sample is going to evolve.
That the path towards these non-ergodic states is complicated
follows from the observation that samples may 'hesitate' for a
long time between the two states, and may evolve in one direction
to end up in the other. In Fig. \ref{fig4}, some of the
'hesitating' samples are indicated in color. It clearly shows how
a few samples in the intermediate concentration region that
behaved consistently like glassy samples before the ergodicity
breaking point, end up as gels at late times. Perhaps even more
surprisingly, Fig. \ref{fig4}b shows that even if one looks at a
single observable such as the non-ergodicity parameter, a
crossover between the two behaviors can be observed; this is most
evident in the data for $1.9 wt \%$. The data shown in the figure
are mere examples; approximately $20 \%$ of the samples  in the
intermediate concentration region behaved in an ambiguous way in
the sense that they seemed to have a hard time to 'decide' whether
they were glasses or gels.

In conclusion, the nature of the non-ergodic state in Laponite suspensions has
been the subject of considerable controversy: both colloidal gel
\cite{Nicolai,Mourchid} and colloidal glass formation
\cite{glass,Bellour} have been invoked and were thought to be
mutually exclusive \cite{Nicolai,glass}. We have shown here that gel and glassy
states of Laponite both exist and are well-defined in the limit of low and high
concentrations.  In gels the main
 cause of
aging is the building up of a network. Slowing down of particle
motion will happen when the particles are trapped in the network,
which may be formed already for very low particle concentrations.
On the other hand, the slowing down of motion in glasses happens
due to topological constraints \cite{HS,structural arrest} that
hinder the motion of particles. Therefore they are usually formed
in samples of higher concentrations.  We here provided clear experimental evidence
for how the distinction of gel and glassy states can be made
experimentally.

This allows for the observation that for intermediate concentrations, the transition to
non-ergodicity can occur in either direction (gel or glass), and may be accompanied by 'hesitations' between
 the two directions. A qualitative
explanation for this behavior is provided by the free energy
landscape picture of slow dynamics. Here, one attributes the slow
structural relaxation to the complex pathways that connect the
configurational states on the multidimensional potential surface.
For Laponite our data suggest that there are at least two
metastable minima in the free energy corresponding to gel and
glass states, and that different pathways towards these
non-equilibrium states exist, providing the first evidence for the
existence of such a complicated free-energy landscape. This
phenomenon should be general for colloidal
 systems with attractive interactions between the particles; indeed the recent discovery of "attractive glasses" for
 spherical colloids \cite{attractive glass} also suggests that gels and glasses are not necessarily clearly distinct
 states of matter, but rather metastable minima in an otherwise complicated free-energy landscape, resulting from both steric
 and attractive interactions, as suggested by some simulations \cite{sciortino} and experiments \cite{gel2}.

 \textbf{Acknowledgments}

The research has been supported by FOM organization in
Netherlands. LPS de l'ENS is UMR8550 of the CNRS, associated with
the universities Paris 6 and 7. We would like to thank H. Tanaka
and S. Rafai and T. Niewenhuizen for helpful discussions.

\end {document}